\documentclass{article}
\usepackage{spconf,amsmath,graphicx}

\usepackage{url,amsfonts,amssymb,bm}

\usepackage{enumitem}
\usepackage{algorithm}
\usepackage{algorithmicx, algpseudocode }
\usepackage{multirow}
\usepackage{color}
\usepackage{mathtools}
\usepackage{caption}
\usepackage{subcaption}
\usepackage{amsthm}

\usepackage[colorlinks=true, linkcolor=blue, citecolor=blue, urlcolor=blue]{hyperref}
\hypersetup{pdfcreator  = {LaTeX with hyperref package},
               pdfproducer = {dvips + ps2pdf} }

\usepackage{tikz}

\usetikzlibrary{calc}

\usepackage{tikz,spconf,amsmath,graphicx,bbm}

\usetikzlibrary{positioning}

\newtheorem{remark}{Remark}

\newtheorem{theorem}{Theorem}

\newtheorem{proposition}{Proposition}

\theoremstyle{definition}

\newtheorem{definition}{Definition}

\DeclareMathOperator*{\diag}{diag}

\usepackage{bm}

\long\def\comment#1{}

\newfont{\bbb}{msbm10 scaled 700}

\newfont{\bb}{msbm10 scaled 1100}

\newcommand{\av}{{\bf a}}
\newcommand{\bv}{{\bf b}}

\newcommand{\dv}{{\bf d}}

\newcommand{\fv}{{\bf f}}

\newcommand{\uv}{{\bf u}}

\newcommand{\vv}{{\bf v}}
\newcommand{\xv}{{\bf x}}
\newcommand{\yv}{{\bf y}}
\newcommand{\zv}{{\bf z}}

\newcommand{\Am}{{\bf A}}

\newcommand{\Cm}{{\bf C}}
\newcommand{\Dm}{{\bf D}}

\newcommand{\Gm}{{\bf G}}
\newcommand{\Hm}{{\bf H}}
\newcommand{\Id}{{\bf I}}
\newcommand{\Jm}{{\bf J}}

\newcommand{\Lm}{{\bf L}}
\newcommand{\Mm}{{\bf M}}

\newcommand{\Qm}{{\bf Q}}

\newcommand{\Sm}{{\bf S}}
\newcommand{\Tm}{{\bf T}}
\newcommand{\Um}{{\bf U}}
\newcommand{\Wm}{{\bf W}}
\newcommand{\Vm}{{\bf V}}

\newcommand{\Zm}{{\bf Z}}
\newcommand{\Lam}{{\bf \Lambda}}

\newcommand{\Ac}{{\cal A}}
\newcommand{\Bc}{{\cal B}}

\newcommand{\Ec}{{\cal E}}

\newcommand{\Gc}{{\cal G}}

\newcommand{\Vc}{{\cal V}}

\newcommand{\Lcb}{{\bm {\mathcal L}}}

\newcommand{\Lambdam}{\hbox{\boldmath$\Lambda$}}

\title{ Spectral folding and two-channel filter-banks on arbitrary graphs }
\name{Eduardo Pavez$^\star$, Benjamin Girault${}^\#{}^\star$, Antonio Ortega$^\star$,  Philip A. Chou$^\dagger$ \thanks{ Author email: pavezcar@usc.edu.    This work was funded  by a Google Faculty Research Award.}}
\address{$^\star$University of Southern California, Los Angeles, California, USA \\
$^\#$ Université de Rennes, ENSAI, CNRS, CREST-UMR 9194, Rennes, FRANCE \\
$^\dagger$Google Research, Seattle, Washington, USA }
\begin{document}
\ninept
\maketitle
\begin{abstract}
 In the past decade, several  multi-resolution representation theories for graph signals have been proposed. Bipartite filter-banks stand out as the most natural extension of time domain filter-banks, in part because perfect reconstruction,  orthogonality and bi-orthogonality conditions in the graph spectral domain resemble those for traditional filter-banks. Therefore,  many of the well known orthogonal and bi-orthogonal designs can be easily adapted for graph signals. A major limitation is that  this framework can only be applied to the normalized Laplacian of bipartite graphs. In this paper we  extend this theory to arbitrary graphs and positive semi-definite variation operators. Our approach is based on a different definition of the graph Fourier transform (GFT), where orthogonality is defined with the respect to the $\Qm$ inner product. We  construct  GFTs satisfying a  spectral folding property, which allows us to easily construct orthogonal and bi-orthogonal perfect reconstruction filter-banks. 
We illustrate signal representation and computational efficiency  of our  filter-banks on  3D point clouds with hundreds of thousands of points.
\end{abstract}
\section{Introduction}
%
%
 Graph signal processing (GSP) provides a toolbox for analysis  and manipulation of  signals living in irregular  domains \cite{ortega2018graph,shuman2013emerging}.  
Given the success of  multi-resolution representations (MRR)  to  analyze and process    traditional signals \cite{vetterli1995wavelets}, significant efforts have been put into extending them  for  graph signals   \cite{shuman2020localized}.

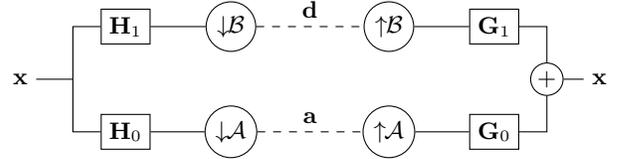
\begin{figure}[t]
    \centering
\def\hpdeltay{1}
\def\lpdeltay{-1}
\def\inputtosplitx{1}
\def\splittofilterx{1}
\def\filtertosamplingx{2}
\def\minspacedownupx{3}
\def\filtertooutputsplitx{1}
\begin{tikzpicture}[scale=0.7]
    \node (input) at (0, 0) {$\xv$};
    \coordinate (input_split) at (\inputtosplitx,0);

	\node[draw,rectangle] (H1) at ($(\inputtosplitx+\splittofilterx,\hpdeltay)$) {$\Hm_1$};
	\node[draw,circle,inner sep=2pt] (downB) at ($(H1)+(\filtertosamplingx,0)$) {$\downarrow\mkern-6mu \Bc$};

	\node[draw,rectangle] (H0) at ($(\inputtosplitx+\splittofilterx,\lpdeltay)$) {$\Hm_0$};
	\node[draw,circle,inner sep=2pt] (downA) at ($(H0)+(\filtertosamplingx,0)$) {$\downarrow\mkern-6mu\Ac$};

	\node[draw,circle,inner sep=2pt] (upB) at ($(downB)+(\minspacedownupx,0)$) {$\uparrow\mkern-6mu\Bc$};
	\node[draw,rectangle] (G1) at ($(upB)+(\filtertosamplingx,0)$) {$\Gm_1$};

	\node[draw,circle,inner sep=2pt] (upA) at ($(downA)+(\minspacedownupx,0)$) {$\uparrow\mkern-6mu\Ac$};
	\node[draw,rectangle] (G0) at ($(upA)+(\filtertosamplingx,0)$) {$\Gm_0$};

    \node[draw,circle, inner sep=1pt](add) at ($(G1)+(\filtertooutputsplitx,\lpdeltay)$) {$+$};


    \node (output) at ($(add)+(\inputtosplitx,0)$) {$\xv$};

    \draw (input) -- (input_split);
	\draw (input_split) |- (H1) -- (downB);
	\draw (input_split) |- (H0) -- (downA);

	\draw[dashed] (downB) -- node[above] {$\dv$} (upB);
	\draw[dashed] (downA) -- node[above] {$\av$} (upA);

    \draw (upB) -- (G1) -| (add);
    \draw (upA) -- (G0) -| (add);

	\draw (add) -- (output);
\end{tikzpicture}
    \caption{Perfect reconstruction two channel filter-bank with analysis filters $\Hm_0$, $\Hm_1$,  and synthesis filters  $\Gm_0$ and $\Gm_1$. $\av$ and $\dv$ denote approximation (low pass) and detail (high pass) coefficients respectively.  Sampling sets are denoted by $\Ac$ and $\Bc$.  }
    \vspace{-0.65cm}
    \label{fig:2chanfb}
\end{figure}

Applications often require  these MRRs   to: (i) be perfect reconstruction (invertible), (ii) be critically sampled (non redundant),  (iii) be orthogonal, and  (iv) have compact support (polynomial filter implementation).  
In the graph setting, it has proven challenging to find theories that satisfy more than a few of these properties simultaneously.  Current theories  require strong assumptions on the graph topology (e.g., bipartite\cite{narang2012perfect,narang2013compact}, circulant \cite{kotzagiannidis2019splines}), and are valid for a single type of graph  operator (e.g., normalized Laplacian or adjacency).
%
Narang and Ortega \cite{narang2012perfect,narang2013compact} proposed two channel filter-banks on bipartite graphs, composed of   graph  filters,  vertex down-sampling, and vertex up-sampling    operators (see Figure \ref{fig:2chanfb}). 
These bipartite filter-banks (BFB) obey (i), (ii), and either (iii) or (iv),   can be designed in the frequency domain, and can be implemented using low degree polynomials. In addition, regular domain filter-banks   can be  easily converted to the graph domain \cite{narang2012perfect,narang2013compact,sakiyama2016spectral,tay2015techniques,tay2017bipartite}.  
Given   their  strong  theoretical properties and   efficient implementations,  BFB have found  numerous applications \cite{anis2016compression,tzamarias2019compression,levorato2012reduced,qiao2019target}.  

Despite all these remarkable properties,  BFB theory \cite{narang2012perfect,narang2013compact}     only applies to   normalized Laplacians and  adjacency matrices of bipartite graphs. These are major limitations  since the graph structure is rarely bipartite (which dictates the down-sampling operator),  whereas the graph  variation operator (or graph shift) is determined by the application. 
%
%
To overcome these issues,  we propose a new  theory that can be applied to: 1) arbitrary graphs, 2) any vertex partition for down-sampling, and 3) positive semi-definite variation operators   (see \cite{girault2018irregularity,anis2016efficient,egilmez2017graph,pavez2020ragft} for examples). The proposed filter-banks  also satisfy (i),  (ii), and either (iii) or (iv), as with BFB.

The BFB theory   is built upon  a {\em spectral folding } property  satisfied by the eigenvectors and eigenvalues of the normalized Laplacian  of bipartite graphs. 
Our theory follows a similar strategy, by proving a  new {\em spectral folding property}  for the $(\Mm,\Qm)$ graph Fourier transform ($(\Mm,\Qm)$-GFT). This result builds upon a generalization of  the  graph Fourier transform to arbitrary finite dimensional Hilbert spaces \cite{girault2018irregularity,girault2020graph} with inner product $\langle \xv, \yv \rangle_{\Qm} = \yv^{\top} \Qm \xv$ and variation operator $\Mm$. 
%
In our framework, the down-sampling and variation operators $\Mm$ completely determine the choice of inner product $\Qm$.   
Interestingly, our spectral domain conditions on the filters match those of  \cite{narang2012perfect,narang2013compact} for the normalized Laplacian of bipartite graphs, and therefore we can re-use any of their filter designs, or any of the more recent improvements \cite{sakiyama2016spectral,tay2015techniques}. 
When the graph is bipartite, and  $\Mm$ is  the normalized or combinatorial Laplacian, we recover the nonZeroDC and ZeroDC filter-banks, respectively \cite{narang2012perfect,narang2013compact}.
%
%
%
%

%
 Early MRRs on arbitrary graphs were constructed by scaling and shifting  spectral graph filters  \cite{crovella2003graph,coifman2006diffusion,hammond2011wavelets}. These methods are difficult to invert (e.g., requiring least squares), are not critically sampled, and lack orthogonality. More recent approaches are redundant \cite{shuman2015multiscale,sakiyama2014oversampled},  lack perfect reconstruction \cite{li2019scalable,anis2017critical}, or change the graph to a bipartite one \cite{narang2010local,zeng2017bipartite,jiang2019admm,narang2012perfect}. While  some of  these approaches \cite{narang2010local,zeng2017bipartite,narang2012perfect}  can exploit efficient filter-bank implementations, once a sparse bipartite graph is available,   obtaining the bipartite graph itself, either by graph  approximation or through graph learning may be computationally infeasible for large graphs. 
%
 More recently,  \cite{sakiyama2019two} proposed graph filter-banks with  spectral domain down-sampling. This sampling operator induces spectral folding of the GFT which is  exploited to obtain perfect reconstruction conditions. Although this approach can be used for arbitrary graphs and  variation operators, it requires computing a full GFT,  which does not scale well to large graphs.
 In contrast to previous approaches, we show that the  proposed filter-banks can be implemented efficiently on large graphs (e.g., with hundreds of thousands of nodes), as long as these are sparse, and outperform BFB in energy compaction and run time.
%

 The rest of the paper is organized as follows. In Sections \ref{sec_gsp_MQ} and \ref{sec_prel_2chanfb} we review the fundamentals of GSP on arbitrary Hilbert spaces, and two channel filter-banks on bipartite graphs, respectively. Our theory  is presented in Section \ref{sec_folding}.  We end this paper with  numerical results and conclusions in  Sections \ref{sec_exp} and  \ref{sec_conclusion}, respectively.
\section{GSP in arbitrary Hilbert spaces}
\label{sec_gsp_MQ}
Scalars, vectors and matrices are written in lower case regular, lower case bold and upper case bold respectively (e.g.,  $a$, $\bv$, $\Cm$). Positive definite and semi-definite matrices are denoted by $\Am \succ 0$ and $\Am \succeq 0$ respectively. 
%
Consider a weighted undirected graph $\Gc = (\Vc, \Ec, \Mm)$ with vertex set $\Vc = \lbrace 1,2,\cdots,n \rbrace $, edge set $\Ec \subset \Vc \times \Vc$, and variation operator $\Mm = (m_{ij})$, satisfying  $m_{ij} = m_{ji}\neq 0$ when $ij \in \Ec$, and $m_{ij}=0$ otherwise.  A graph signal is a function $x: \Vc \rightarrow \mathbb{R}$, that can be represented by a vector $\xv = [x_1,\cdots,x_n]^{\top}$. 
The variation operator is assumed to be positive semi-definite, and the variation of a signal is   $ \Delta(\xv) = \xv^{\top} \Mm \xv \geq 0$.
Intuitively, signals with increased variation are said to have higher frequency content.  We will further assume that $\Mm$ is irregular, that is, the graph is connected.
Typical examples of variation operators include the combinatorial and normalized Laplacian matrices. For a symmetric non negative matrix $\Wm = (w_{ij})$,  degree of node $i$ is $d_i = \sum_{j \in \Vc }w_{ij}$, and the degree matrix is $\Dm = \diag(d_1,\cdots, d_n)$. The combinatorial Laplacian is $\Lm = \Dm - \Wm$, while the normalized Laplacian is $\Lcb = \Dm^{-1/2}\Lm \Dm^{-1/2} = \Id - \Dm^{-1/2} \Wm \Dm^{-1/2}$. 
 \cite{girault2018irregularity}  introduced the idea of using an inner product $ \langle \xv, \yv \rangle_{\Qm} = \yv^{\top} \Qm \xv$, and  induced norm given by $\Vert \xv \Vert_{\Qm} = \sqrt{\langle \xv, \xv \rangle_{\Qm}}$,
with $\Qm \succ 0$.
The $(\Mm,\Qm)$-GFT basis vectors are the columns of  $\Um = [\uv_1,\cdots, \uv_n]$, which solve the  generalized eigenvalue problem
\begin{equation}
    \Mm \uv_k = \lambda_k \Qm \uv_k, 
\end{equation}
and  $0 \leq \lambda_1 \leq ,\cdots, \leq \lambda_n$. The set of eigenvalues (spectrum) of a graph is denoted by $\sigma(\Mm,\Qm)$.
The generalized eigenvectors are  $\Qm$-orthonormal,  hence
$ \Vert \uv_i \Vert_{\Qm}   = 1,\quad \forall i \in \Vc$,  and 
 $\langle \uv_i, \uv_j \rangle_{\Qm} = 0,\quad \forall i \neq j$, that is, 
  $\Um^{\top} \Qm \Um = \Id$ in matrix form.
A  graph signal $\xv$ has the following repesentation in the $(\Mm,\Qm)$-GFT basis
\begin{equation}
    \xv = \sum_{i=1}^n \langle \xv, \uv_i \rangle_{\Qm} \uv_i = \Um \hat{\xv}.
\end{equation}
The $(\Mm,\Qm)$-GFT of $\xv$ is denoted by $\hat{\xv}$, with coordinates $\hat{x}_i = \langle \xv, \uv_i \rangle_{\Qm}$. In matrix form this  corresponds to $\hat{\xv} = \Um^{\top} \Qm \xv$, while the inverse transform is given by $\xv = \Um \hat{\xv}$, since $\Um \Um^{\top} \Qm = \Id$.
A linear operator $\Hm$ is a spectral filter  if there is a function $h:\mathbb{R}_+ \rightarrow\mathbb{R}$ so that $  \Hm = \Um h(\Lam) \Um^{\top} \Qm = h(\Zm)$, 
where $\Zm = \Qm^{-1} \Mm = \Um \Lam \Um^{\top} \Qm$ is the fundamental matrix, and $\Lam = \diag(\lambda_1, \cdots, \lambda_n)$.
%
%
\section{Two channel filter-banks}
\label{sec_prel_2chanfb}
In this section we define two channel filter-banks on arbitrary graphs, and review the BFB theory \cite{narang2012perfect,narang2013compact}.
A two channel filter-bank is depicted in Figure \ref{fig:2chanfb}. The analysis  filters are $\Hm_0$ and $\Hm_1$, while the synthesis filters correspond to $\Gm_0$ and $\Gm_1$. Consider the set $\Ac$ and $\Bc = \Ac \setminus \Vc$, which form a partition of the vertex set $\Vc$. Without loss of generality we assume that $\Ac = \lbrace 1, 2,\cdots, \vert \Ac \vert \rbrace$.  Down-sampling a signal $\xv$ on a set $\Ac$ corresponds to keeping the entries $x_i: i \in \Ac$, and discarding the rest. This can be represented by $    \xv_\Ac = \Sm_\Ac \xv$, where $\Sm_\Ac = [\Id_\Ac, \mathbf{0}]$ is a $\vert \Ac \vert \times \vert \Vc \vert$ selection matrix. 
The up-sampling operator is $\Sm_\Ac^{\top}$. Down-sampling followed by up-sampling  sets to zero the entries in $\Bc$, thus $\Sm_\Ac^{\top} \xv_{\Ac} = \Sm_\Ac^{\top} \Sm_\Ac \xv = \begin{bmatrix}  \xv_\Ac^{\top}  & \mathbf{0}     \end{bmatrix}^{\top}$.
\subsection{Vertex domain conditions for arbitrary graphs}
The analysis operator (filtering and down-sampling) from Fig. \ref{fig:2chanfb} is
\begin{equation}
\Tm_a =  \Sm_\Ac^{\top} \Sm_\Ac \Hm_0 + \Sm_\Bc^{\top} \Sm_\Bc \Hm_1 = \begin{bmatrix}
\Sm_\Ac \Hm_0 \\ \Sm_\Bc \Hm_1
\end{bmatrix}.
\end{equation}
The outputs of the low pass and high pass channels, called approximation $\av$ and detail $\dv$ coefficients, respectively, are given by: 
\begin{equation}
\Tm_a \xv = \begin{bmatrix}
\av \\
\dv
\end{bmatrix}
= \begin{bmatrix}
\Sm_\Ac \Hm_0\xv \\ \Sm_\Bc \Hm_1 \xv 
\end{bmatrix}.
\end{equation}
The synthesis operator has a similar expression
\begin{equation}\label{eq_synthesis_operator}
    \Tm_s = \Gm_0 \Sm_\Ac^{\top} \Sm_\Ac  + \Gm_1 \Sm_\Bc^{\top} \Sm_\Bc = \begin{bmatrix}
    \Gm_0 \Sm_\Ac^{\top} & \Gm_1 \Sm_\Bc^{\top} 
    \end{bmatrix}.
\end{equation}
We say that a two channel filter-bank is {\bf perfect reconstruction (PR)}  if   $\Tm_s \Tm_a = \Tm_a \Tm_s = \Id$.
A linear operator $\Tm$ is $\Qm$-{\bf orthogonal}  if  for each $\xv$, the norm of the transformed signal is preserved, that is,  $\Vert \Tm \xv \Vert_{\Qm} = \Vert  \xv \Vert_{\Qm}$. In matrix form this corresponds to $\Tm^{\top} \Qm \Tm = \Qm.$
 For a  PR two channel filter-bank, $\Tm_a$ is  $\Qm$-orthogonal if and only if  $\Tm_s$ is  $\Qm$-orthogonal. 
 Finding operators $\Tm_a$, $\Tm_s$ that are orthogonal and PR is not that difficult, in fact, any non-singular orthogonal matrix  can be used for $\Tm_a$, and the synthesis operator can be chosen as $\Tm_s = \Qm^{-1}\Tm_a^{\top}\Qm$. The  challenge is finding operators  that exploit the graph structure, and that can be  efficiently implemented on large arbitrary graphs. In the next subsection we review the approach of \cite{narang2012perfect,narang2013compact} to design BFB using spectral graph filters. 
\subsection{Spectral domain conditions for bipartite graphs}
BFBs can be constructed on bipartite graphs:
\begin{definition}
A graph $\Gc=(\Vc,\Ec)$ is   bipartite on  $(\Ac,\Bc)$, if 
i)     $(\Ac,\Bc)$ forms a partition, that is, $\Ac \cap \Bc = \emptyset$, and $\Ac \cup \Bc = \Vc$, 
and ii) for all $(i,j)\in \Ec$, $i \in \Ac$ and $j \in \Bc$, or $i \in \Bc$ and $j \in \Ac$.
\end{definition}
In bipartite graphs, only edges between sets $\Ac$ and $\Bc$ are allowed, therefore  the Laplacian matrices have the  form
\begin{equation}
    \Lm = \begin{bmatrix}
    \Dm_\Ac & -\mathbf{W}_{ \Ac\Bc} \\ -\mathbf{W}_{\Bc\Ac} & \Dm_\Bc
    \end{bmatrix}, \quad 
    \Lcb =  \begin{bmatrix}
    \mathbf{I}_\Ac & -\tilde{\mathbf{W}}_{\Ac\Bc} \\ -\tilde{\mathbf{W}}_{\Bc\Ac} & \mathbf{I}_\Bc
    \end{bmatrix},
\end{equation}
where $\Lcb = \Dm^{-1/2}\Lm\Dm^{-1/2} = \mathbf{I} - \tilde{\mathbf{W}}$, and  $\tilde{\mathbf{W}} = \Dm^{-1/2}\Wm\Dm^{-1/2}$.
 Spectral filters are defined using the $(\Lcb,\Id)$-GFT,  thus $\Zm = \Lcb$, and 
\begin{equation}\label{eq_spec_graf_filt_normL}
  \Hm_i = h_i(\Lcb),  \quad \Gm_i = g_i(\Lcb), \quad \textnormal{ for }i \in \{ 0,1 \}.
\end{equation} 
We define $\Jm = \diag(\fv)$, where
\begin{equation}\label{eq_f_setindicator}
    \fv_{i} = \left\{ \begin{array}{cc}
		1  & \mbox{if } i \in \Ac \\
		-1 & \mbox{if } i \in \Bc.
	\end{array} \right.
\end{equation}
Then we can write the  operators as $
 \Sm_\Ac^{\top} \Sm_\Ac = \frac{1}{2}(\Id +\Jm)$, and $\Sm_\Bc^{\top} \Sm_\Bc = \frac{1}{2}(\Id -\Jm)$.
The PR condition now becomes
\begin{equation}\label{eq_pr_vertex_equation}
  \Id 
  =\frac{1}{2}\left( \Gm_0 \Hm_0 + \Gm_1 \Hm_1\right) + \frac{1}{2}\left( \Gm_0 \Jm \Hm_0 - \Gm_1 \Jm \Hm_1\right).
\end{equation}
The BFB framework attains PR by designing filters that obey
\begin{equation}\label{eq_alias_cancellation}
    \Gm_0 \Hm_0 + \Gm_1 \Hm_1 = 2\Id,\quad \textnormal{ and }\quad
     \Gm_0 \Jm \Hm_0 - \Gm_1 \Jm \Hm_1 = \mathbf{0}.
\end{equation}
\begin{theorem}\cite{narang2012perfect}\label{th_PR_bipartite}
For a BFB with filters given by (\ref{eq_spec_graf_filt_normL}), a necessary and sufficient condition for PR is  that $ \forall \lambda \in \sigma(\Lcb,\Id)$, 
\begin{align}\label{eq_pr1}
    h_0(\lambda)g_0(\lambda) +  h_1(\lambda)g_1(\lambda) &= 2\\
    h_0(\lambda)g_0(2-\lambda) - h_1(\lambda)g_1(2-\lambda) &= 0. \label{eq_pr2}
\end{align}
\end{theorem}
Theorem \ref{th_PR_bipartite} is proven   using the following property. 
\begin{proposition}\cite{narang2012perfect}[Spectral folding.]
\label{prop_folding_bipartite}
If $\Lcb \uv = \lambda \uv $, then $\Jm \uv$ is also an eigenvector with eigenvalue $2 - \lambda$.
\end{proposition}
  $\Id$-orthogonal  filter-banks  can be realized if  and only if  for all $\lambda \in\sigma(\Lcb,\Id)$, the filters also satisfy
\begin{align}\label{eq_orthogonal_filters1}
    h_0^2(\lambda) + h_1^2(\lambda) &= 2,\\
    h_0(\lambda)h_0(2-\lambda) - h_1(\lambda)h_1(2-\lambda) &= 0.\label{eq_orthogonal_filters2}
\end{align}
Orthogonal filter-banks are  PR, while the converse is not true in general. In fact,   filters $h_0, h_1, g_0, g_1$ obey (\ref{eq_orthogonal_filters1}) and (\ref{eq_orthogonal_filters2}), if and only if,  they obey (\ref{eq_pr1}), (\ref{eq_pr2}) and $h_i = g_i$.
Orthogonal filter-banks do not have polynomial implementations, thus requiring full eigendecomposition for implementation. A popular approach  to overcome this is to approximate the filters with Chebyshev polynomials \cite{hammond2011wavelets,sakiyama2016spectral}.
An  alternative is to use perfect reconstruction  bi-orthogonal filters, 
\begin{equation}\label{eq_bior}
h_0(\lambda) = g_1(2-\lambda), \quad 
h_1(\lambda) = g_0(2-\lambda), 
\end{equation}
which   can be designed to be near orthogonal and polynomial \cite{narang2013compact}.
The proofs of orthogonality and PR  require  Proposition \ref{prop_folding_bipartite}, which holds only for the normalized Laplacian of bipartite graphs. To the best of our knowledge no other variation operators have this property for bipartite or arbitrary graphs. 
\section{Main results}
\label{sec_folding}
In this  section we   extend Proposition \ref{prop_folding_bipartite} and the BFB theory to arbitrary graphs and positive semi-definite variation operators by using an alternative definition of GFT.  
\begin{theorem}[Spectral folding]
\label{prop:folding-variation}
Consider an arbitrary partition of the vertices, $\Ac$ and $\Bc = \Vc \setminus \Ac$. Without loss of generality we assume $\Ac = \{ 1,2,\cdots,\vert \Ac \vert \}$.
Let $\Mm \succeq 0$ be a  variation operator, and  
\begin{equation}\label{eq_M_Q}
    \Mm = \begin{bmatrix}
    \Mm_{\Ac\Ac} & \Mm_{\Ac\Bc} \\ \Mm_{\Bc\Ac} & \Mm_{\Bc\Bc}
    \end{bmatrix}, \quad \Qm=\begin{bmatrix}
    \Mm_{\Ac\Ac} & \mathbf{0} \\ \mathbf{0} & \Mm_{\Bc\Bc}
    \end{bmatrix}
\end{equation}
where $\Qm$ is the inner product matrix. If  $\Qm \succ 0$, then
 \begin{equation}
     \Mm \uv = \lambda \Qm \uv \iff \Mm \Jm\uv = (2-\lambda) \Qm \Jm\uv.
 \end{equation}
\end{theorem}
 We sketch the proof of ($\Rightarrow$), since the other direction follows from the same argument. 
Let $\uv = [\uv_\Ac^{\top}, \uv_\Bc^{\top} ]^{\top}$ be a generalized eigenvector of $\Mm$ with eigenvalue $\lambda$, then
\begin{equation*}
    \Mm_{\Ac\Ac} \uv_\Ac + \Mm_{\Ac\Bc} \uv_\Bc = \lambda \Qm_\Ac \uv_\Ac,\;
    \Mm_{\Bc\Ac} \uv_\Ac + \Mm_{\Bc\Bc} \uv_\Bc = \lambda \Qm_\Bc \uv_\Bc.
\end{equation*}
Set $\mathbf{v} = \Jm\uv=[\uv_\Ac^{\top}, -\uv_\Bc^{\top} ]^{\top}$, and compute $\Mm \vv$. A simple calculation and using the fact that   $\uv$ is a generalized eigenvector, and   $\Mm_{\Ac\Ac} = \Qm_\Ac$, and $\Mm_{\Bc\Bc} = \Qm_\Bc$ produces the desired result.
For this $(\Mm,\Qm)$-GFT, the fundamental matrix is
\begin{equation*}
    \Zm = \Qm^{-1}\Mm = \begin{bmatrix}
    \Id_{\Ac} & \Mm_{\Ac\Ac}^{-1} \Mm_{\Ac \Bc} \\
    \Mm_{\Bc\Bc}^{-1} \Mm_{\Bc\Ac} &\Id_{\Bc}
    \end{bmatrix}
= \Um \Lambdam \Um^{\top}\Qm,
\end{equation*}
 and  $\Lambdam = \diag(\lambda_1, \lambda_2,\cdots,\lambda_n)$.
%
%
This $(\Mm,\Qm)$-GFT shares some properties with the $(\Lcb,\Id)$-GFT of bipartite graphs.  
First, the generalized eigenvalues obey $0 \leq \lambda_i \leq 2$, and inequalities become strict when $\Mm$ is non-singular.
Second,  the  eigenvalue $\lambda = 1$ has multiplicity at least $\vert \vert \Ac \vert - \vert \Bc \vert \vert$,  thus the middle graph frequency is less selective when the down-sampling sets have uneven size.
Finally,   when $\Mm = \Lm + \Vm$ and  $\Vm$ is diagonal, i.e., $\Lm$ is a generalized Laplacian, the multiplicity of $\lambda_1$ (smallest eigenvalue) is equal to the number of connected components of the graph.
%
%
Now we state conditions for PR filter-banks.
\begin{theorem}[Perfect reconstruction]\label{th_PR_arbitrary}
For any positive semi definite variation operator $\Mm$, and any vertex partition  $\Vc = \Ac \cup \Bc$. Choose the inner product matrix $\Qm$ according to Theorem \ref{prop:folding-variation}, and spectral graph filters for $i \in \lbrace 0,1 \rbrace$
 \begin{equation}\label{eq_spec_graf_filt_Z}
  \Hm_i =  \Um h_i(\Lam)\Um^{\top}\Qm, \quad
  \Gm_i = \Um g_i(\Lam)\Um^{\top}\Qm.
\end{equation}
The functions $h_i, g_i$ for $i \in \{0,1 \}$ obey conditions (\ref{eq_pr1}) and (\ref{eq_pr2}) for all $\lambda \in \sigma(\Mm,\Qm)$, if and only if  the   filter-bank of Figure \ref{fig:2chanfb} is PR.
\end{theorem}
 %
%
The proof follows from  Theorem \ref{prop:folding-variation} and similar arguments as those used in \cite{narang2012perfect} to prove that (\ref{eq_alias_cancellation}) is true.
Theorem \ref{th_PR_arbitrary} implies that our framework can be implemented using  filters designed for bipartite graphs (see \cite{narang2012perfect,narang2013compact,tay2015techniques,sakiyama2016spectral}). 
We can also construct  $\Qm$-orthogonal filter-banks. 
\begin{theorem}[Parseval]\label{th_orthogonality_arbitrary}
Under the  conditions of Theorem \ref{th_PR_arbitrary},  the analysis filters obey (\ref{eq_orthogonal_filters1}) and (\ref{eq_orthogonal_filters2}) for all $\lambda \in \sigma(\Mm,\Qm)$, if and only if, 
\begin{equation*}
 \langle \Tm_a \xv, \Tm_a \yv \rangle_{\Qm}   =  \langle \xv, \yv \rangle_{\Qm}, \quad
 \langle \Tm_s \xv, \Tm_s \yv \rangle_{\Qm}   =  \langle \xv, \yv \rangle_{\Qm}, \forall \xv, \yv.
\end{equation*}
\end{theorem}
In particular, we have preservation of the $\Qm$ norm, thus $  \Vert \Tm_a \xv \Vert^2_{\Qm}   = \Vert  \xv \Vert^2_{\Qm}$, and $ \Vert \Tm_s \xv \Vert^2_{\Qm}   = \Vert  \xv \Vert^2_{\Qm}$.
 When $\Qm = \Id$, the synthesis operator is  the transpose of $\Tm_a$, however, in general  we have the relation
 \begin{equation}
     \Tm_s = \Qm^{-1} \Tm_a^{\top} \Qm.
 \end{equation}
%
%
It was demonstrated in \cite{narang2013compact} that orthogonal filter-banks cannot be implemented with polynomials of $\Lcb$. The same arguments can be used to show that the proposed orthogonal filter-banks cannot be implemented as polynomials of $\Zm$.   As an alternative, \cite{narang2013compact} proposed the bi-orthogonal filters (\ref{eq_bior}), which have polynomials implementations.   
Although these filters are not orthogonal, we can use  the  reasoning from [Section III-B]\cite{narang2013compact}, to show the analysis and synthesis operators can be designed to be approximately $\Qm$-orthogonal, and satisfy
\begin{equation}
    \alpha  \Vert \xv \Vert_{\Qm} \lesssim \Vert \Tm_{*}\xv \Vert_{\Qm} \lesssim \beta \Vert \xv \Vert_{\Qm}\;\; \forall \xv,
\end{equation}
where $*$ can be $a$ or $s$, and 
\begin{equation*}
    \alpha^2 =\frac{1}{2} \inf_{\lambda \in [0,2]} (h_0^{2}(\lambda) + h_1^{2}(\lambda)), \; 
    \beta^2 =\frac{1}{2} \sup_{\lambda \in [0,2]} (h_0^{2}(\lambda) + h_1^{2}(\lambda)).
\end{equation*}
\begin{remark} The zero-DC filter-bank was  proposed in  \cite{narang2013compact}, so that  the smoothest  basis function is constant. This approach can be implemented by multiplying the input signal by $\Dm^{1/2}$ before applying the analysis filter-bank, and multiplying by $\Dm^{-1/2}$ at the output of the synthesis filter-bank. This ensures that a constant input signal has zero response in the high pass channel. \cite{narang2013compact} showed that bi-orthogonal zero-DC filter-banks can be implemented with polynomials of the  random walk Laplacian of a bipartite graph. The zero-DC filter-banks can be derived as a special case of our framework, by noticing that for bipartite graphs with Laplacian $\Lm$, the inner product matrix from Proposition \ref{prop:folding-variation}  is  $\Qm = \Dm$, and the fundamental matrix is the random walk Laplacian $\Zm = \Dm^{-1} \Lm$. 
\end{remark}
\begin{figure}[h]
    \centering
    \includegraphics[width=0.45\textwidth]{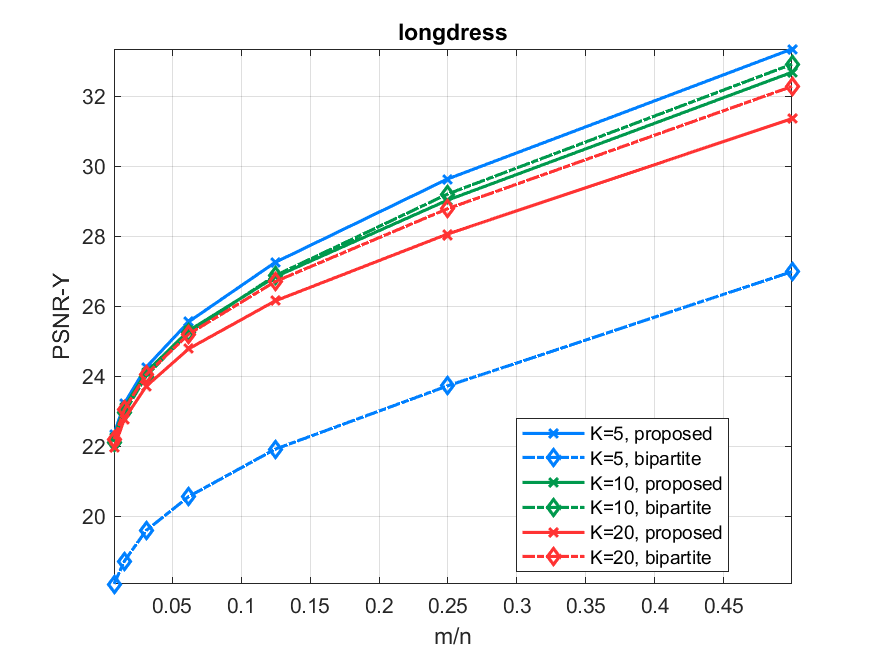}
    \caption{Linear approximation of 3D point clouds attributes using iterated two channel filter-banks.}
    \label{fig:psnr}
\end{figure}
\begin{figure}[ht]
    \centering
    \includegraphics[width=0.45\textwidth]{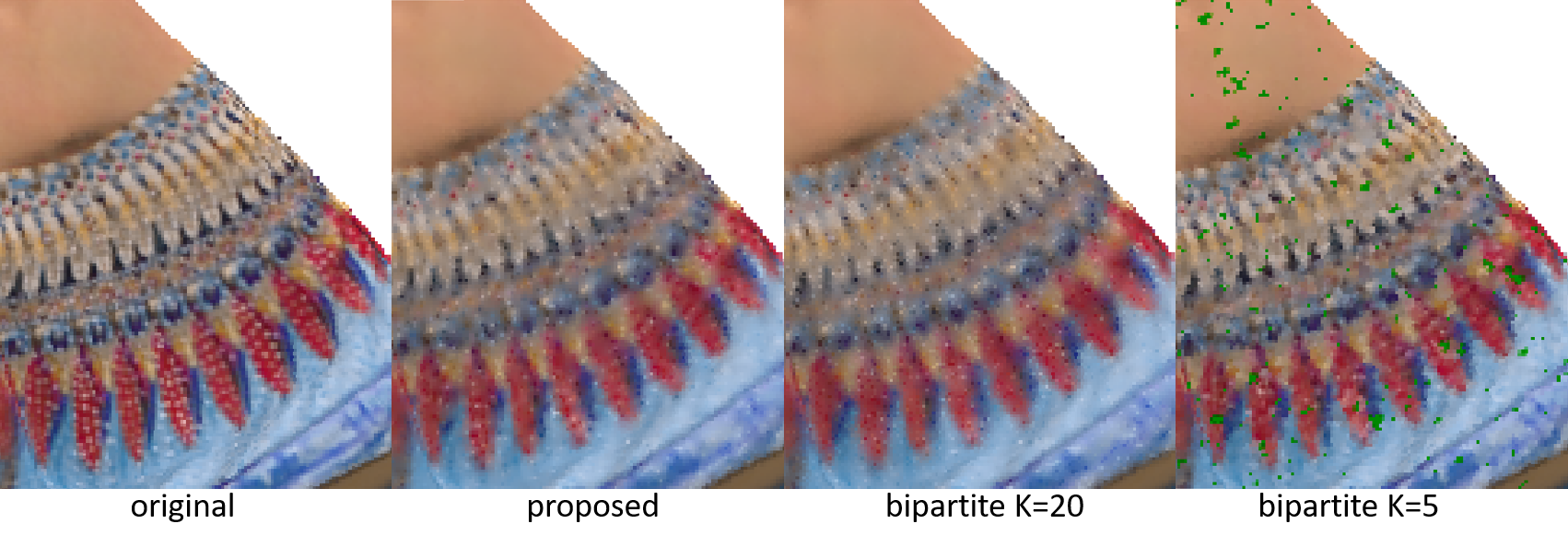}
    
    \includegraphics[width=0.45\textwidth]{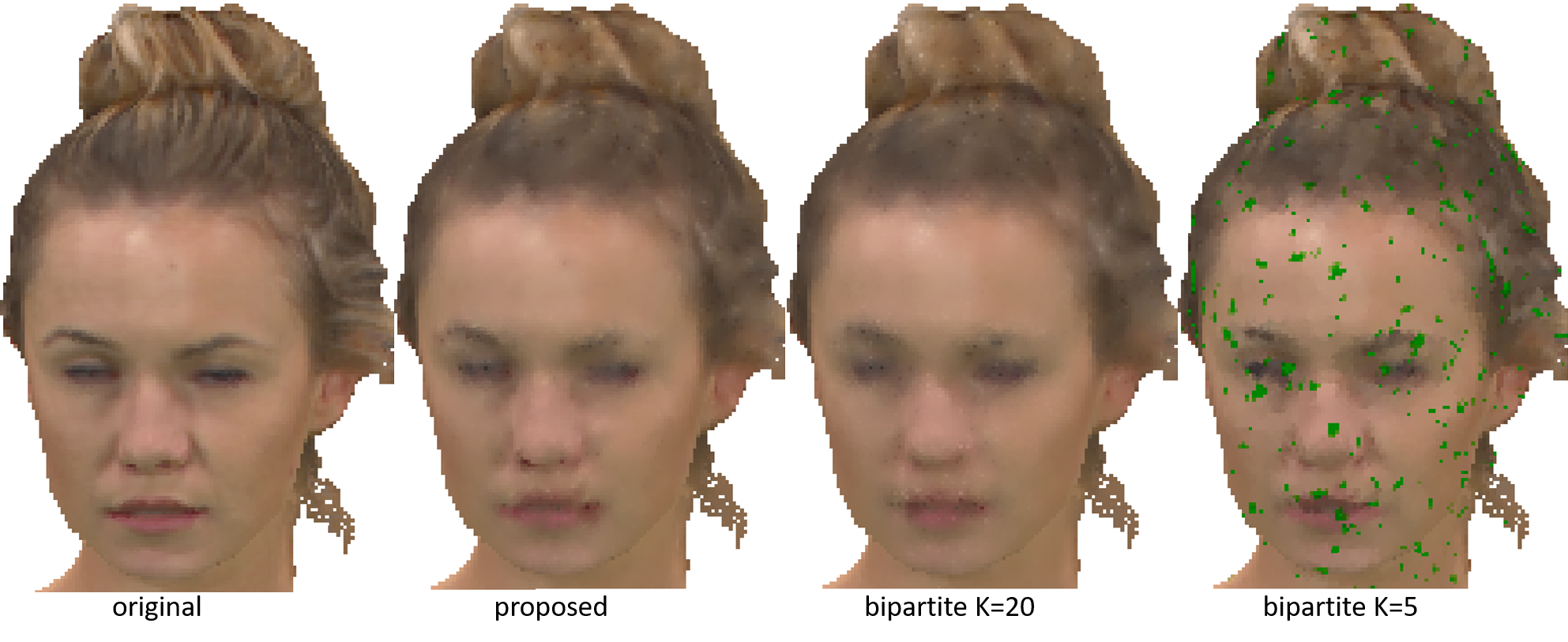}
    \caption{Top $m/n = 0.25$, bottom $m/n = 0.125$. Comparison of linear approximation using BFB and proposed filter-banks.}
    \label{fig:visual}
\end{figure}
\section{Numerical results}
\label{sec_exp}
%
We consider the ``lazy" bi-orthogonal filter-bank, with filters $\Hm_0 = \Id$, $\Hm_1 = \Zm$, $\Gm_0 = 2\Id - \Zm$, and $\Gm_1 = \Id$. The down-sampling set $\Ac$ is chosen so that a node $i \in \Ac$ with probability $1/2$,  leading to $\vert \Ac \vert \approx \vert \Bc \vert$. We apply this  filter-bank iteratively to the low pass channel. After $L$ levels of decomposition, we have $L$ high frequency channels, and $1$ low pass channel. The low pass channel has approximately $n 2^{-L}$ coefficients, while the high pass channels at iteration $\ell$ has approximately $n 2^{-\ell}$ coefficients. 

 3D point clouds   consist of a list of points in 3D space represented by their coordinates $ \vv_i = [x_i, y_i, z_i] $, and attributes $\av_i$. We use the ``longdress'' point cloud from the ``8iVFBv2'' dataset \cite{d20178i}, which comes with color attributes. 

 Each point $\vv_i$ can be assigned  a node in a graph,  while  an edge between nodes $i$ and $j$ is  added if  $i$ is one of the $K$ nearest neighbors of $j$, or vice versa. Edge weights are computed as $w_{ij} = 1/\Vert \vv_i - \vv_j  \Vert $. Vertex down-sampling of point clouds corresponds to selecting a subset of points, therefore we can repeat this graph construction procedure at the output of the low pass channel. To implement the iterated filterbank on 3D point clouds we follow the steps: 1)  construct a graph with $K$ nearest neighbor (KNN) algorithm and compute  its combinatorial Laplacian matrix, 2) generate of a random down-sampling set $\Ac$, and $\Bc = \Ac^c$, and 3) if a bipartite graph is desired,  keep edges between $\Ac$ and $\Bc$ and remove the rest.
\paragraph*{Signal representation.}
  We apply the iterated filter-bank described above with $L=7$ levels to the color attributes of a single frame of the ``longdress'' sequence. For each $L$, we  reconstruct the color signal using only the low frequency coefficients. We compare the proposed filter-banks with the BFBs as a function of $m/n$, where $m$ is the number of coefficients in the low pass channel, and $n$ is the total number of coefficients.  Figure \ref{fig:psnr} shows  that the proposed filter-bank has better energy compaction when the KNN graph has fewer edges ($K=5$), and performance decreases as  $K$ increases. The best performance of the BFBs is achieved with an intermediate value of $K=10$.   In Figure \ref{fig:visual} we show the reconstructions of  regions of the point cloud. When the bipartite graph is too sparse ($K=5$), several artifacts can be observed, which can be attributed to points/nodes in the high pass channel that do not have connections in the low pass channel. When the bipartite graph is denser $K=20$, the reconstruction has no artifacts, however details are smoothed more aggressively. The proposed filter-bank, with a sparser graph,  better preserves textures and  facial features (e.g., eyes, mouth and hair).
\paragraph*{Complexity.}
We compare the run time of the iterated analysis filter-bank ($L=7$) applied to $20$ frames of the ``longdress'' sequence. We run the experiment using Matlab on a desktop computer.  The bipartite filter-bank with $K=20$ and $K=10$ takes $14.3$ and $8.23$ seconds per frame respectively, while the proposed filter-bank with the best performance ($K=5$),   takes $6.72$ seconds per frame. These point clouds have an average of $795,000$ points per frame. The complexity of our implementation is dominated by two factors. Graph construction, which is implemented using approximate KNN, with complexity proportional to $K$. Complexity of filtering is dominated by the product $\Zm \xv$. When the graph is bipartite, $\Zm \xv$ is a sparse matrix-vector product. In the non-bipartite case, $\Zm\xv$ is computed in two steps, first we perform a sparse matrix-vector product $ \yv = \Lm \xv$,  and then solve the linear system $\Qm \zv = \yv$. Since $\Qm$ is sparse, a numerically accurate approximation of $\zv = \Qm^{-1}\yv$ can be found efficiently using the  ``\textbackslash'' operator in Matlab. 
%
%
%
%
\section{Conclusion}
\label{sec_conclusion}
This paper extended the graph filter-bank framework  \cite{narang2012perfect,narang2013compact}, which uses the normalized Laplacian of bipartite graphs, to positive semi-definite variation operators, and arbitrary graphs. We achieve this by proving that the spectral folding property  is not unique to the normalized Laplacian of bipartite graphs, and in fact, it is satisfied by certain generalized eigenvalues and eigenvectors of non-bipartite graphs. Based on this, we proposed a new graph Fourier transform that is orthogonal in a $\Qm$ inner product, and that leads to perfect reconstruction, orthogonal and biorthogonal conditions, equivalent to those already known for the bipartite graph case. We implemented a simple degree $1$ polynomial filter-bank on  3D point clouds graphs  hundreds of thousands of vertices. Our numerical results indicate that our framework outperform the same filter-bank, implemented with bipartite graphs at various sparsity levels.  The proposed filter-bank outperforms bipartite filter-banks in   run-time and energy compaction. 
\vfill
\pagebreak
\bibliographystyle{IEEEbib}
\bibliography{refs}

\end{document}